\documentclass[aps,preprint,floatfix,nofootinbib,showpacs]{revtex4-1}
\pdfoutput=1
\usepackage{graphicx,color,float}
\usepackage{hyperref}

\begin{document}

{\small
\begin{flushright}
IUEP-HEP-18-01
\end{flushright} }

\title{
Measuring properties of a Heavy Higgs boson\\[0mm] in the 
$H\to t\bar t \to bW^+\bar b W^-$ decay
}

\def\slash#1{#1\!\!/}
\def\lsim{\:\raisebox{-0.5ex}{$\stackrel{\textstyle<}{\sim}$}\:}
\def\gsim{\:\raisebox{-0.5ex}{$\stackrel{\textstyle>}{\sim}$}\:}
\newcommand{\imag}{\Im {\rm m}}
\newcommand{\real}{\Re {\rm e}}

\renewcommand{\thefootnote}{\arabic{footnote}}

\author{
Jung Chang$^{1,2}$, Kingman Cheung$^{2,3,4}$, Jae Sik Lee$^{1,2,5}$,
Chih-Ting Lu$^4$, and  Jubin Park$^{5,1,2}$}
\affiliation{
$^1$ Department of Physics, Chonnam National University, 
300 Yongbong-dong, Buk-gu, Gwangju, 500-757, Republic of Korea \\
$^2$ Physics Division, National Center for Theoretical Sciences,
Hsinchu, Taiwan \\
$^3$ Division of Quantum Phases and Devices, School of Physics, 
Konkuk University, Seoul 143-701, Republic of Korea \\
$^4$ Department of Physics, National Tsing Hua University,
Hsinchu 300, Taiwan \\
$^5$ Institute for Universe and Elementary Particles, Chonnam National University, \\
300 Yongbong-dong, Buk-gu, Gwangju, 500-757, Republic of Korea 
}
\date{March 14, 2018}

\begin{abstract}
Suppose a heavy neutral Higgs or scalar boson $H$ is 
discovered at the LHC, it is important to investigate its couplings 
to the standard model particles as much  as possible.
Here in this work we attempt to probe the CP-even and CP-odd couplings
of the heavy Higgs boson to a pair of top quarks, 
through the decay $H \to t\bar t \to b W^+ \bar b W^-$. 
We use the helicity-amplitude method to
write down the most general form for the angular distributions
of the final-state $b$ quarks and $W$ bosons.
We figure out that there are 6 types of angular observables and,
under CP$\widetilde{\rm T}$ conservation, 
one-dimensional angular distributions can only reveal two of them.
Nevertheless, the $H$ couplings to the
 $t\bar t$ pair can be fully determined 
by exploiting the one-dimensional angular distributions. 
A Higgs-boson mass of 380 GeV not too far above the $t\bar t$ threshold is
illustrated with full details.  With a total of $10^4$ events of $H
\to t\bar t \to bW^+ \bar b W^+$, one can determine the couplings up to 
10-20\% uncertainties.
\end{abstract}

\maketitle

\section{Introduction}
The scalar boson that was discovered at the LHC in
2012~\cite{atlas,cms} turned out to be best described by the Standard
Model (SM) Higgs boson~\cite{Cheung:2013kla}, which is remarkable
confirmation of the Higgs mechanism proposed in 1960s~\cite{higgs}.
Among the Higgs boson couplings to the SM particles,
the most constrained one is its coupling to the massive gauge bosons
that is very close to the corresponding SM value with about 10\%
uncertainty \cite{Cheung:2014noa}. Nevertheless, the couplings to
fermions are much less constrained, especially for the first two
generations. The coupling to the top-quark pair from global fits
has about 20--30\% uncertainty~\cite{Cheung:2014noa}. There was also
direct measurements of the top Yukawa coupling in $pp \to t\bar t h$
production \cite{top-yukawa}, which still needs more data to have 
more precise measurements  than the global fitting.

%
%
%

Even though the SM has achieved a great success in accounting for the 
interactions among the basic building blocks of matter,
however extra particles and new interactions are required to explain
the experimental observations of dark matter, non-vanishing neutrino mass, 
the baryon asymmetry of our Universe, inflation, etc.
In most extensions beyond the SM, the Higgs sector is enlarged to
include more than one Higgs doublet resulting in charged Higgs bosons
and several neutral Higgs bosons in addition to the one discovered at
the LHC.
For example, the minimal supersymmetric extension of the 
SM, aka MSSM~\cite{HPN}, requires two Higgs doublet fields, thus leading 
to a pair of charged Higgs bosons and 3 neutral ones. 
In the next-to-minimal supersymmetric standard model, there are 
two additional neutral Higgs bosons~\cite{Cheung:2010ba}.
%

Suppose that in future experiments a neutral Higgs boson $H$
heavier than the SM 125 GeV Higgs boson (denoted by $h$) is discovered.
It is generally expected that the decays of the heavy Higgs boson $H$ 
into gauge bosons would be suppressed 
as it becomes heavier and heavier, 
because the measurement of the 
gauge-Higgs coupling of the 
SM-like 125 GeV boson allows only 
a little room for the $H$ couplings to massive vector bosons.
However, its couplings to fermions have no reasons to be small. 
Indeed, once it is above
the $t\bar t$ threshold,
the decay into $t\bar t$ pair could be dominant.

In this work, we assume that the heavy Higgs boson $H$ is not too far above
the $t\bar t$ threshold, say 380 GeV, and the dominant decay mode is $t\bar t$.
Without requiring $H$ to carry any definite CP-parity,
we consider the possibility to probe its CP nature via the decay
$H \to t\bar t \to b W^+ \bar b W^-$.
We employ the helicity-amplitude method \cite{Hagiwara:1985yu} to 
calculate the decay amplitude taking into account all spin
correlations in the decay chain. By measuring various angular distributions,
in particular the angle between the decay planes of the top and anti-top
quarks, one can discern the CP-even and CP-odd couplings of the 
Higgs boson. This is the main goal of this work.
Other fermionic modes, in general, are either too small or suffer
tremendously from SM backgrounds.  The top quark also 
has the advantage that it decays before hadronization, in contrast
to the bottom or charm quarks, and therefore the spin information is 
retained in the decay products of the top quark. Thus, the spin and CP
properties of the parent Higgs boson can be determined 
by studying several kinematical distributions of
the decay products of the top and anti-top quarks.

On the other hand, 
when the heavy Higgs boson $H$ is below the $t\bar t$ threshold,
its bosonic decay modes $ZZ$, $hh$, and $hZ$ are
useful for probing the CP nature of it.
By taking account of the spin-$0$ nature of $H$, 
only the $ZZ$ mode may
lead to nontrivial angular correlations among the decay products of
the $Z$ bosons through the interferences among various helicity states
of the two intermediate $Z$ bosons before their decays \cite{higgs-zz}.
This bosonic mode was suggested to determine 
the spin and parity of the Higgs boson~\cite{Choi:2002jk} quite 
a number of years ago.
%
%
After the 125 GeV Higgs-boson discovery, the method was practically applied to
determine the spin and CP properties of 
the observed Higgs boson~\cite{Bolognesi:2012mm,Modak:2013sb}.
We shall not concern with the bosonic modes in this work.

Under the current experimental status, in which active 
searches for heavy resonances decaying into a $t\bar t$ pair 
have been continually performed~\cite{h_tt},
our study may show how well one can determine
the properties of such a heavy scalar Higgs boson
at the LHC and/or High Luminosity LHC (HL-LHC).
We refer to, for example, Ref.~\cite{Grzadkowski:1995rx}
for some previous studies at $e^+ e^-$ or $\mu^+ \mu^-$
colliders.

The remainder of this article is organized as follows. In Sec.~II,
based on the helicity amplitude method~\cite{Hagiwara:1985yu},
we present a formalism for the study of angular distributions
in the decay $H\to t \bar t \to b W^+ \bar b W^-$.
We point out that there are 6 types of
angular observables in general 
and we can classify them according to 
the CP and CP$\widetilde{\rm T}$ parities of each observable.
In Sec.~III, we illustrate how well one can measure the couplings of 
the heavy Higgs boson by exploiting the angular observables
introduced in Sec.~II.
Finally, Sec.~IV is devoted to a brief summary,
some prospects for future work and conclusions.

\section{Formalism}

Without loss of generality, the Lagrangian describing
the interactions of the Higgs boson $H$ with top quarks can be written as
\cite{Lee:2003nta}
\begin{equation}
{\cal L}_{H\bar t t}=-g_t H\,\bar t (g^S+i\gamma_5 g^P) t
=-g_t\sum_{A=L,R} (g^S+iA g^P)\,  H\, \bar t P_A t
\end{equation}
where $P_A=(1+A\gamma_5)/2$ with $A=-(L)\,,+(R)$. $g_t$ is the overall 
strength of
the $H$--$t$--$\bar t$ coupling and $g^{P(S)}=0$ when $H$ is the
pure 
CP-even (odd) state.
If $H$ is a CP-mixed state, both $g^S$ and $g^P$ are non-vanishing.
For the SM Higgs boson, $g_t = g m_t/(2M_W)$, $g^S=1$, and $g^P=0$.
On the other hand,
the Lagrangian describing the interactions of the top quarks
with bottom quarks and $W$ bosons is
\begin{eqnarray}
{\cal L} &=& -\frac{g}{\sqrt{2}}\left[
W_\mu^- \bar b \gamma^\mu (f_L\,P_L+f_R\,P_R)t \ + \
W_\mu^+ \bar t \gamma^\mu (f_L^*\,P_L+f_R^*\,P_R)b \right] \nonumber \\[3mm]
         &=& -\frac{g}{\sqrt{2}}\left[
(W_\mu^+)^* \bar b \gamma^\mu (f_L\,P_L+f_R\,P_R)t \ + \
(W_\mu^-)^* \bar t \gamma^\mu (f_L^*\,P_L+f_R^*\,P_R)b \right]\,.
\end{eqnarray}
In the SM, $f_L=1$ and $f_R=0$.

\subsection{Helicity Amplitudes}
We first present the helicity amplitude for the process 
$$ H\to t(p_t,\sigma_t) \bar{t}(\bar p_t,\bar\sigma_t)
\to b(p_b,\sigma_b) W^+(k_1,\epsilon_1)\,
\bar{b}(\bar p_b,\bar\sigma_b) W^-(k_2,\epsilon_2).$$
Here $p_{t,b}$ and $\bar p_{t,b}$ are
four-momenta of the quarks ($t,b$) and antiquarks ($\bar t,\bar b$), respectively,
and $\sigma_{t,b}$ and $\bar \sigma_{t,b}$ denote their helicities.
The four-momenta of $W^+$ and $W^-$ are denoted by $k_1$ and $k_2$, respectively,
with $p_t=p_b+k_1$ and $\bar p_t=\bar p_b+k_2$ and 
$\epsilon_1(\lambda_1)$ and $\epsilon_2(\lambda_2)$ are 
the polarization 4-vectors of $W$ bosons.
Depending on the helicities of the final-state particles,
the amplitude can be cast into the form
\begin{eqnarray}
i{\cal M}_{\sigma_b\lambda_1:\bar\sigma_b\lambda_2}&=&
\sum_{A,B,C=L,R}\Bigg\{
\bar u_b(p_b,\sigma_b)\left[-i\frac{g}{\sqrt{2}}\slash\epsilon_1^*(k_1)f_AP_A\right]
\frac{i}{\slash p_t-m_t}\left[-ig_t(g^S+iBg^P)P_B\right]
\nonumber \\[2mm]
&&\times
\frac{i}{-\slash{\bar p}_t-m_t}
\left[-i\frac{g}{\sqrt{2}}\slash\epsilon_2^*(k_2)f_C^*P_C\right] v_b(\bar
p_b,\bar\sigma_b) \Bigg\}
\nonumber \\[2mm]
&=& -i\frac{1}{p_t^2-m_t^2+im_t\Gamma_t}\,\frac{1}{\bar p_t^2-m_t^2+im_t\Gamma_t}
\sum_{\sigma_t,\bar\sigma_t}
{\cal M}^{t\to bW^+}_{\sigma_t:\sigma_b\lambda_1}
{\cal M}^{H\to t\bar t}_{\sigma_t\bar\sigma_t}
{\cal M}^{\bar t\to \bar bW^-}_{\bar\sigma_t:\bar\sigma_b\lambda_2}
\end{eqnarray}
using
\begin{equation}
\sum_{\sigma_t} u(p_t,\sigma_t)\bar u(p_t,\sigma_t) = \slash{p}_t +m_t\,, \ \ \
\sum_{\bar\sigma_t} v(\bar p_t,\bar\sigma_t)
\bar v(\bar p_t,\bar \sigma_t) = \bar \slash{p}_t -m_t\,.
\end{equation}

The helicity amplitude for the decay 
$H(q)\to t(p_t,\sigma_t) \bar t(\bar p_t,\bar\sigma_t)$
in the rest frame of $H$ is given by
\begin{equation}
{\cal M}^{H\to t \bar t}_{\sigma_t\bar\sigma_t}
=g_t\sqrt{s}\langle\sigma_t\rangle\delta_{\sigma_t\bar\sigma_t}\,
e^{-i\sigma_t\phi_t}
\end{equation}
where $s=q^2$, $\phi_t$ is the azimuthal angle of the $t$ momentum, and
\begin{eqnarray}
\label{eq:httamp}
\langle+\rangle &=& X\, g^S -i\, Y\, g^P \,, \ \ \
\langle-\rangle = -X\, g^S -i\, Y\, g^P\,.
\end{eqnarray}
The momentum-dependent $X$ and $Y$ are given by
\begin{eqnarray}
X&=&
\frac{\left[(1+\lambda_H^{1/2})^2-(\alpha_t-\bar\alpha_t)^2\right]^{1/2}
-\left[(1-\lambda_H^{1/2})^2-(\alpha_t-\bar\alpha_t)^2\right]^{1/2}}{2}\,,
\nonumber \\[2mm]
Y&=&
\frac{\left[(1+\lambda_H^{1/2})^2-(\alpha_t-\bar\alpha_t)^2\right]^{1/2}
+\left[(1-\lambda_H^{1/2})^2-(\alpha_t-\bar\alpha_t)^2\right]^{1/2}}{2}\,,
\end{eqnarray}
where $\lambda_H=(1+\alpha_t^2+\bar\alpha_t^2
- 2\alpha_t-2\bar\alpha_t-2\alpha_t\bar\alpha_t)$ 
with $\alpha_{t} = p_{t}^2/s$ and $\bar\alpha_t = \bar p_t^2/s$.
When the top quarks are on-shell, $X=\beta_t=(1-4m_t^2/s)^{1/2}$ and $Y=1$.
One may take $\phi_t=0$ without loss of generality.

The helicity amplitude for the decay 
$t(p_t,\sigma_t) \to b(p_b,\sigma_b) W^+(k_1,\epsilon_1)$
in the $t$ rest frame is given by
\begin{equation}
{\cal M}^{t\to b W^+}_{\sigma_t:\sigma_b \lambda} 
=
-\frac{g}{\sqrt{2}}\sqrt{2\sqrt{p_t^2}E_b}\,\langle\sigma_t:\sigma_b\lambda_1\rangle_t\,,
\end{equation}
where $2\sqrt{p_t^2}E_b = p_t^2+m_b^2-M_W^2$.
The reduced helicity amplitudes $\langle\sigma_t:\sigma_b\lambda_1\rangle_t$
are given by
\begin{eqnarray}
\label{eq:tamp}
\langle\sigma_t:\sigma_b\lambda_1\rangle_t =
\left\{ \begin{array}{ll}
\sum_{A=L,R}\left[-A(\sqrt{2}f_Ac_{\theta_1/2})
\frac{(1+A\sigma_t \beta_b)^{1/2}}{\sqrt{2}}
\delta_{\sigma_t\lambda_1} \delta_{\sigma_t\sigma_b}
\right.  &  \\[4mm]
\left. \hspace{1.55cm} +
A\sigma_t(\sqrt{2}f_As_{\theta_1/2}e^{+i\sigma_t\phi_1})
\frac{(1-A\sigma_t \beta_b)^{1/2}}{\sqrt{2}}
\delta_{\sigma_t-\lambda_1} \delta_{\sigma_t-\sigma_b} \right]
& {\rm for}~\lambda_1=\pm \\[6mm]
\sum_{A=L,R}\left[(f_As_{\theta_1/2})
\frac{\beta_bE_b+A\sigma_tE_W}{M_W}
\frac{(1+A\sigma_t \beta_b)^{1/2}}{\sqrt{2}} \delta_{\sigma_t\sigma_b}
\right.  &  \\[4mm]
\left. \hspace{1.55cm} +
\sigma_t(f_Ac_{\theta_1/2}e^{+i\sigma_t\phi_1})
\frac{\beta_bE_b-A\sigma_tE_W}{M_W}
\frac{(1-A\sigma_t \beta_b)^{1/2}}{\sqrt{2}} \delta_{\sigma_t-\sigma_b} \right]
& {\rm for}~\lambda_1=0
\end{array} \right.
\end{eqnarray}
where $\theta_1$ and $\phi_1$ are the polar and azimuthal angles of
the $W^+$ momentum in the $t$ rest frame and
$c_{\theta_1/2}=\cos(\theta_1/2)$ and
$s_{\theta_1/2}=\sin(\theta_1/2)$.
We note that $\sigma_b=\lambda_1$ when $\lambda_1=\pm$ and
the 4 amplitudes of
$\langle\pm:+-\rangle_t$ and 
$\langle\pm:-+\rangle_t$ are identically vanishing.
In the $m_b\to 0$ limit, the reduced amplitudes simplify
and we have the following non-vanishing 8 amplitudes
\footnote{
In Ref.~\cite{Kane:1991bg}, the authors presented the helicity amplitudes
in the $m_b\to 0$ limit.  We find 
a minor discrepancy in four of the amplitudes with $\lambda_1=\pm$
by an overall factor of $e^{-i\lambda_1\phi_1}$, which does not affect
the full amplitude squared
for the process ${H\to t\bar t \to bW^+ \bar b W^-}$.}:
\begin{eqnarray}
&&
\langle-:--\rangle_t =+\sqrt{2}f_Lc_{\theta_1/2}\,, \ \ \ \hspace{0.9cm}
\langle-:-0\rangle_t =+\frac{\sqrt{p_t^2}}{M_W}f_Ls_{\theta_1/2}\,, \ \ \ \nonumber \\[2mm]
&&
\langle+:--\rangle_t =-\sqrt{2}f_Ls_{\theta_1/2}e^{+i\phi_1}\,, \ \ \
\langle+:-0\rangle_t =+\frac{\sqrt{p_t^2}}{M_W}f_Lc_{\theta_1/2}e^{+i\phi_1}\,; \nonumber \\[4mm]
&&
\langle+:++\rangle_t =-\sqrt{2}f_Rc_{\theta_1/2}\,, \ \ \ \hspace{0.9cm}
\langle+:+0\rangle_t =+\frac{\sqrt{p_t^2}}{M_W}f_Rs_{\theta_1/2}\,, \ \ \ \nonumber \\[2mm]
&&
\langle-:++\rangle_t =-\sqrt{2}f_Rs_{\theta_1/2}e^{-i\phi_1}\,, \ \ \
\langle-:+0\rangle_t =-\frac{\sqrt{p_t^2}}{M_W}f_Rc_{\theta_1/2}e^{-i\phi_1}\,.
\end{eqnarray}
We note that $A=\sigma_b$  in the $m_b=0$ limit.

The helicity amplitude for the decay 
$\bar t(\bar p_t,\bar \sigma_t) \to
\bar b(\bar p_b,\bar \sigma_b) W^-(k_2,\epsilon_2)$
in the $\bar t$ rest frame is similarly given by
\begin{equation}
{\cal M}^{\bar t\to \bar b W^-}_{\bar\sigma_t:\bar\sigma_b \lambda_2}  \equiv
-\frac{g}{\sqrt{2}}\sqrt{2\sqrt{\bar p_t^2}\bar E_b}\,
\langle\bar\sigma_t:\bar\sigma_b\lambda_2\rangle_{\bar t}
\end{equation}
where $2\sqrt{\bar p_t^2}\bar E_b = \bar p_t^2+m_b^2-M_W^2$ and
the reduced amplitudes $\langle\bar\sigma_t:\bar\sigma_b\lambda_2\rangle_{\bar t}$
can be obtained by replacing 
$f_A$ with $f_{-A}^*$ in Eq.~(\ref{eq:tamp})
together with $\sigma_{t,b} \to \bar\sigma_{t,b}$, $\lambda_1\to \lambda_2$, etc
\footnote{For details of the
relation between the helicity amplitudes for the $t$ and $\bar t$ decays, 
see Appendix A.}.
Further we note the relations
\begin{eqnarray}
\langle-:--\rangle_{\bar t} &=& -\langle+:++\rangle_t^* \,, \ \ \
\langle-:-0\rangle_{\bar t}  =  +\langle+:+0\rangle_t^* \,, \nonumber \\ [2mm]
\langle+:--\rangle_{\bar t} &=& +\langle-:++\rangle_t^* \,, \ \ \
\langle+:-0\rangle_{\bar t}  =  -\langle-:+0\rangle_t^* \,, \nonumber \\ [2mm]
\langle+:++\rangle_{\bar t} &=& -\langle-:--\rangle_t^* \,, \ \ \
\langle+:+0\rangle_{\bar t}  =  +\langle-:-0\rangle_t^* \,, \nonumber \\ [2mm]
\langle-:++\rangle_{\bar t} &=& +\langle+:--\rangle_t^* \,, \ \ \
\langle-:+0\rangle_{\bar t}  =  -\langle+:-0\rangle_t^* \,.
\end{eqnarray}

Collecting all the sub-amplitudes we obtain
\begin{eqnarray}
\label{eq:fullamp1}
{\cal M}_{\sigma_b\lambda_1:\bar\sigma_b\lambda_2}&\equiv &
-g^2g_t\,\frac{\sqrt{s}}{2}\,
\frac{\sqrt{p_t^2+m_b^2-M_W^2}}{p_t^2-m_t^2+im_t\Gamma_t}\,
\frac{\sqrt{\bar p_t^2+m_b^2-M_W^2}}{\bar p_t^2-m_t^2+im_t\Gamma_t}
\langle\sigma_b\lambda_1:\bar\sigma_b\lambda_2\rangle
\end{eqnarray}
where 
\begin{equation}
\label{eq:fullamp2}
\langle\sigma_b\lambda_1:\bar\sigma_b\lambda_2\rangle = \sum_{\sigma_t=\pm}
\langle\sigma_t\rangle
\langle\sigma_t:\sigma_b\lambda_1\rangle_t
\langle\sigma_t:\bar\sigma_b\lambda_2\rangle_{\bar t}\,.
\end{equation}

\subsection{Angular coefficients and observables}
The partial decay width of the process $H\to t\bar t \to b W^+ \bar bW^-$
is given by
\footnote{For the four-body phase space, see Appendix B.}
\begin{eqnarray}
\label{eq:partial}
d\Gamma &=& \frac{N_C}{2 M_H} \left(
\sum_{\sigma_b,\lambda_1,\bar\sigma_b,\lambda_2}
\left|{\cal M}_{\sigma_b\lambda_1:\bar\sigma_b\lambda_2}\right|^2
\right)\, d\Phi_4
\nonumber \\[2mm]
&=& \frac{N_C}{2^{13}\, \pi^6\, M_H}\,
\lambda_H^{1/2} \lambda_t^{1/2} \lambda_{\bar t}^{1/2}
\sqrt{p_t^2}\sqrt{\bar p_t^2} \left(\sum|{\cal M}|^2\right)\
d\sqrt{p_t^2}\,d\sqrt{\bar p_t^2}\,dc_{\theta_1}\,dc_{\theta_2}\,d\Phi
\end{eqnarray}
where $N_C=3$ and
\begin{eqnarray}
\lambda_t&=&\left(1-\frac{m_b^2}{p_t^2}-\frac{M_W^2}{p_t^2}\right)^2
-4\frac{m_b^2}{p_t^2}\,\frac{M_W^2}{p_t^2}\,, \ \ \
\lambda_{\bar t}=\left(1-\frac{m_b^2}{\bar p_t^2}-\frac{M_W^2}{\bar p_t^2}\right)^2
-4\frac{m_b^2}{\bar p_t^2}\,\frac{M_W^2}{\bar p_t^2}\,. 
\end{eqnarray}
For any values of $f_L$ and $f_R$, taking also account of finite $m_b$ effects,
the precise differential angular distribution $\frac{{\rm d}\Gamma}
{ {\rm d}c_{\theta_{1}}{\rm d}c_{\theta_{2}} {\rm d}\Phi }$
can be obtained numerically by integrating Eq.~(\ref{eq:partial})
over $\sqrt{p_t^2}$ and $\sqrt{\bar p_t^2}$
and using Eqs.  
(\ref{eq:fullamp1}),
(\ref{eq:fullamp2}),
(\ref{eq:httamp}), and
(\ref{eq:tamp}).

On the other hand, in the $m_b\to 0$ limit, 
the amplitudes take much simpler forms and
one can derive {\it analytic} expressions for the differential angular distributions
in terms of
physically meaningful
angular coefficients and observables.
When $f_L=1$ and $f_R=0$,
there are only four non-vanishing amplitudes:
$\langle--:++\rangle$,
$\langle--:+0\rangle$,
$\langle-0:++\rangle$, and
$\langle-0:+0\rangle$.
\begin{table}[!t]
\caption{\label{tab:smamps}
{\it The non-vanishing amplitudes when $m_b=0$ taking $f_L=1$, and $f_R=0$.
}
}
\begin{center}
\begin{tabular}{ccc|c||ccc|c}
\hline 
$\sigma_t$ & $\sigma_b$ & $\lambda_1$ & 
$\langle\sigma_t:\sigma_b\lambda_1\rangle_t$ &
$\bar\sigma_t$ & $\bar\sigma_b$ & $\lambda_2$ & 
$\langle\bar\sigma_t:\bar\sigma_b\lambda_2\rangle_{\bar t}$ \\[0mm]
\hline
$+$ & $-$ & $-$ & $-\sqrt{2}s_{\theta_1/2}e^{i\phi_1}$ &
$+$ & $+$ & $+$ & $-\sqrt{2}c_{\theta_2/2}$ \\
$+$ & $-$ & $0$ & $\frac{\sqrt{p_t^2}}{M_W}c_{\theta_1/2}e^{i\phi_1}$ &
$+$ & $+$ & $0$ & $\frac{\sqrt{\bar p_t^2}}{M_W}s_{\theta_2/2}$ \\[0mm]
\hline
$-$ & $-$ & $-$ & $\sqrt{2}c_{\theta_1/2}$ &
$-$ & $+$ & $+$ & $-\sqrt{2}s_{\theta_2/2}e^{-i\phi_2}$ \\
$-$ & $-$ & $0$ & $\frac{\sqrt{p_t^2}}{M_W}s_{\theta_1/2}$ &
$-$ & $+$ & $0$ & $-\frac{\sqrt{\bar p_t^2}}{M_W}c_{\theta_2/2}e^{-i\phi_2}$ \\[0mm]
\hline
\end{tabular}
\end{center}
\end{table}
Explicitly, from Table~\ref{tab:smamps}, we have
\begin{eqnarray}
\langle--:++\rangle&=&
2\left(\langle+\rangle s_{\theta_1/2}c_{\theta_2/2}e^{i\phi_1}
-\langle-\rangle c_{\theta_1/2}s_{\theta_2/2}e^{-i\phi_2}\right)
\nonumber \\[2mm]
\langle--:+0\rangle&=&
-\sqrt{2}\frac{\sqrt{\bar p_t^2}}{M_W}
\left(\langle+\rangle s_{\theta_1/2}s_{\theta_2/2}e^{i\phi_1}
+\langle-\rangle c_{\theta_1/2}c_{\theta_2/2}e^{-i\phi_2}\right)
\nonumber \\[2mm]
\langle-0:++\rangle&=&
-\sqrt{2}\frac{\sqrt{p_t^2}}{M_W}
\left(\langle+\rangle c_{\theta_1/2}c_{\theta_2/2}e^{i\phi_1}
+\langle-\rangle s_{\theta_1/2}s_{\theta_2/2}e^{-i\phi_2}\right)
\nonumber \\[2mm]
\langle-0:+0\rangle&=&
\frac{\sqrt{p_t^2}\sqrt{\bar p_t^2}}{M_W^2}
\left(\langle+\rangle c_{\theta_1/2}s_{\theta_2/2}e^{i\phi_1}
-\langle-\rangle s_{\theta_1/2}c_{\theta_2/2}e^{-i\phi_2}\right)
\end{eqnarray}
where $\theta_{1(2)}$ and $\phi_{1(2)}$ denote the direction of $W^{+(-)}$ in 
the $t(\bar t)$ rest frame.
And then, the sum of the amplitudes squared can be organized as
\begin{eqnarray}
\sum_{\sigma_b,\lambda_1,\bar\sigma_1,\lambda_2} 
\left|\langle\sigma_b\lambda_1:\bar\sigma_b\lambda_2\rangle\right|^2 &=&
C_1
\left[ \left(1+\frac{p_t^2\bar p_t^2}{4M_W^4}\right)
\left(1-c_{\theta_1}c_{\theta_2}\right)
+\frac{p_t^2+\bar p_t^2}{2M_W^2}
\left(1+c_{\theta_1}c_{\theta_2}\right)\right]
\nonumber \\[2mm]
&+&
C_2
\left[ \left(-1+\frac{p_t^2\bar p_t^2}{4M_W^4}\right)
\left(c_{\theta_1}-c_{\theta_2}\right)
+\frac{p_t^2-\bar p_t^2}{2M_W^2}
\left(c_{\theta_1}+c_{\theta_2}\right)\right]
\nonumber \\[2mm]
&+& 
C_3\left(\frac{p_t^2}{2M_W^2}-1\right) \left(\frac{\bar p_t^2}{2M_W^2}-1\right)
(-s_{\theta_1}s_{\theta_2}c_\Phi)\nonumber \\[2mm]
&+&
C_4\left(\frac{p_t^2}{2M_W^2}-1\right) \left(\frac{\bar p_t^2}{2M_W^2}-1\right)
s_{\theta_1}s_{\theta_2}s_\Phi 
\end{eqnarray}
with $\Phi=\phi_1+\phi_2$ denoting the angle between the two decay planes and
the 4 angular coefficients are given by
\begin{eqnarray}
C_1 &\equiv & \left|\langle+\rangle\right|^2+\left|\langle-\rangle\right|^2
= 2\left[|X|^2\left(g^S\right)^2+|Y|^2\left(g^P\right)^2\right]\,, 
\nonumber \\[2mm]
C_2 &\equiv & \left|\langle+\rangle\right|^2-\left|\langle-\rangle\right|^2
= 4\,\imag(X^*Y)\,g^Sg^P\,,
\nonumber \\[2mm]
C_3 &\equiv & 2\,\real\left[\langle+\rangle\langle-\rangle^*\right]
= 2\left[-|X|^2\left(g^S\right)^2+|Y|^2\left(g^P\right)^2\right]\,,
\nonumber \\[2mm]
C_4 &\equiv & 2\,\imag\left[\langle+\rangle\langle-\rangle^*\right]
= 4\,\real(X^*Y)\,g^Sg^P\,.
\end{eqnarray}
Under CP  and CP$\widetilde{\rm T}$~\footnote{
$\widetilde{\rm T}$ denotes the naive  time-reversal transformation
under which the  matrix element gets complex conjugated.}
transformations, the reduced $H$-$t$-$\bar t$ helicity  amplitudes  transform as follows:
\begin{equation}
\langle \pm \rangle
\, \stackrel{\rm CP}{\leftrightarrow}  \,
\langle \mp \rangle\,, \qquad
\langle \pm \rangle
\, \stackrel{\rm CP\widetilde{\rm T}}{\leftrightarrow} \,
\langle \mp \rangle^*\,.
\end{equation}
We note that the CP parities of $C_1$, $C_2$ ,$C_3$ and $C_4$
are $+$, $-$, $+$, and $-$, respectively,
implying that $C_2$ and $C_4$
are non-vanishing  only when $g^S$ and $g^P$
exist simultaneously.
Furthermore, the angular coefficient $C_2$ is 
CP$\widetilde{\rm T}$ odd and it can be induced
only in the presence of non-vanishing absorptive (or imaginary) parts of 
$X$ and $Y$.

By integrating Eq.~(\ref{eq:partial})
over $\sqrt{p_t^2}$ and $\sqrt{\bar p_t^2}$, we have
\begin{equation}
\frac{{\rm d}\Gamma}
{ {\rm d}c_{\theta_{1}}{\rm d}c_{\theta_{2}} {\rm d}\Phi }
=N_C\frac{g^4g_t^2 \sqrt{s}}{2^{15}\pi^6}
\sum_{i}{\cal F}_i f_i(\theta_1,\theta_2,\Phi)
\end{equation}
where $\sqrt{s}=M_H$ and $i=11,12,21,22,3,4$.
The angular functions are
\begin{eqnarray}
f_{11}(\theta_1,\theta_2,\Phi)&=&1-c_{\theta_1}c_{\theta_2}\,,  \ \ \
f_{12}(\theta_1,\theta_2,\Phi) = 1+c_{\theta_1}c_{\theta_2}\,,  \nonumber \\[2mm]
f_{21}(\theta_1,\theta_2,\Phi)&=&c_{\theta_1}-c_{\theta_2}\,,  \ \ \ \ \,
f_{22}(\theta_1,\theta_2,\Phi) = c_{\theta_1}+c_{\theta_2}\,, \nonumber \\[2mm]
f_{3}(\theta_1,\theta_2,\Phi) &=&-s_{\theta_1}s_{\theta_2}c_\Phi\,,  \nonumber \\[2mm]
f_{4}(\theta_1,\theta_2,\Phi) &=& s_{\theta_1}s_{\theta_2}s_\Phi\,,
\end{eqnarray}
and the numerical factors ${\cal F}_i$ are given by
\begin{eqnarray}
\label{eq:calFi}
{\cal F}_i&=&
\int \lambda_H^{1/2}\lambda_t^{1/2}\lambda_{\bar t}^{1/2}
\sqrt{p_t^2}\sqrt{\bar p_t^2} \
\widetilde{C}_i(p_t^2,\bar p_t^2)\
\nonumber \\
&&\hspace{2.5cm} \times
\frac{p_t^2+m_b^2-M_W^2} {(p_t^2-m_t^2)^2+m_t^2\Gamma_t^2}\
\frac{\bar p_t^2+m_b^2-M_W^2} {(\bar p_t^2-m_t^2)^2+m_t^2\Gamma_t^2}\
d\sqrt{p_t^2}\,d\sqrt{\bar p_t^2}
\end{eqnarray}
in which the tilded $6$ angular coefficients $\widetilde{C}_{i}$ are related to
$C_i$ as follows:
\begin{eqnarray}
\widetilde{C}_{11}&=&C_1\left(1+\frac{p_t^2\bar p_t^2}{4M_W^4}\right)\,, \ \ \
\widetilde{C}_{12} = C_1\,\frac{p_t^2 +\bar p_t^2}{2M_W^2}\,, 
\nonumber \\[2mm]
\widetilde{C}_{21}&=&C_2\left(-1+\frac{p_t^2\bar p_t^2}{4M_W^4}\right)\,, \ \ \
\widetilde{C}_{22} = C_2\,\frac{p_t^2 -\bar p_t^2}{2M_W^2}\,, 
\nonumber \\[2mm]
\widetilde{C}_{3}&=&
C_3\left(\frac{p_t^2}{2M_W^2}-1\right) \left(\frac{\bar p_t^2}{2M_W^2}-1\right)\,,
\nonumber \\[2mm]
\widetilde{C}_{4}&=&
C_4\left(\frac{p_t^2}{2M_W^2}-1\right) \left(\frac{\bar p_t^2}{2M_W^2}-1\right)\,.
\end{eqnarray}
To proceed further,
we have introduced weight factor $w_i$'s which are defined by
\begin{equation}
w_i\equiv\frac{{\cal F}_i}{{\cal F}\overline{\widetilde{C}}_i}
\end{equation}
where
\begin{eqnarray}
\label{eq:calf}
{\cal F}&=&
\int \lambda_H^{1/2} \lambda_t^{1/2} \lambda_{\bar t}^{1/2}
\sqrt{p_t^2}\sqrt{\bar p_t^2} \
\frac{p_t^2+m_b^2-M_W^2} {(p_t^2-m_t^2)^2+m_t^2\Gamma_t^2}\
\frac{\bar p_t^2+m_b^2-M_W^2} {(\bar p_t^2-m_t^2)^2+m_t^2\Gamma_t^2}\
d\sqrt{p_t^2}\,d\sqrt{\bar p_t^2}
\end{eqnarray}
and 
\begin{equation}
\overline{\widetilde{C}}_i=\widetilde{C}_i(p_t^2=m_t^2,\bar p_t^2=m_t^2)
\end{equation}
are the constant tilded angular coefficients at the $t$ pole.
Explicitly, 
\begin{eqnarray}
\overline{\widetilde{C}}_{11}&=&
2\left(1+\frac{m_t^4}{4M_W^4}\right)\left[\beta_t^2\left(g_S\right)^2+
\left(g_P\right)^2\right]\,, \ \ \
\overline{\widetilde{C}}_{12}=
2\left(\frac{m_t^2}{M_W^2}\right)\left[\beta_t^2\left(g_S\right)^2+
\left(g_P\right)^2\right]\,, \nonumber \\[2mm]
\overline{\widetilde{C}}_{21,22}&=&0\,, \nonumber \\[2mm]
\overline{\widetilde{C}}_3&=&
2\left(\frac{m_t^2}{2M_W^2}-1\right)^2\left[-\beta_t^2\left(g_S\right)^2+
\left(g_P\right)^2\right]\,, \ \ \ 
\overline{\widetilde{C}}_4=
4\left(\frac{m_t^2}{2M_W^2}-1\right)^2\beta_tg^Sg^P\,.
\end{eqnarray}
We observe $\overline{\widetilde{C}}_{21,22}$ are identically vanishing because
$X=\beta_t$ and $Y=1$ are real when $p_t^2=\bar p_t^2=m_t^2$.
We also note that
\begin{equation}
\label{eq:c1}
\overline{\widetilde{C}}_{11}+\overline{\widetilde{C}}_{12}
=2\left(1+\frac{m_t^2}{2M_W^2}\right)^2\left[\beta_t^2\left(g_S\right)^2+
\left(g_P\right)^2\right]\,.
\end{equation}

Finally, we have obtained the normalized differential angular distribution
\begin{equation}
\label{eq:approx3d}
\frac{1}{\Gamma}\frac{{\rm d}\Gamma}
{ {\rm d}c_{\theta_{1}}{\rm d}c_{\theta_{2}} {\rm d}\Phi }
=\sum_{i}
\overline{\widetilde{R}}_i
\left( \frac{f_i(\theta_1,\theta_2,\Phi)}{8\pi}\right)
\end{equation}
with the $6$ angular observables defined by
\begin{equation}
\overline{\widetilde{R}}_i \equiv
\omega_i\overline{\widetilde{C}}_i/
(\omega_{11}\overline{\widetilde{C}}_{11}+\omega_{12}\overline{\widetilde{C}}_{12})\,.
\end{equation}
After integrating over any two of the angles $\theta_1$, 
$\theta_2$, and $\Phi$, one can obtain the following
one-dimensional angular distributions 
in terms of the constant $t$-pole angular observables 
$\overline{\widetilde{R}}_i$'s:
\begin{eqnarray}
\label{eq:approx}
\frac{1}{\Gamma}\frac{{\rm d}\Gamma}{{\rm d}c_{\theta_{1}}} &=&
\frac{1}{2}+ \frac{1}{2}\left(\overline{\widetilde{R}}_{21}
+\overline{\widetilde{R}}_{22}\right)c_{\theta_1}\,, \ \ \
\frac{1}{\Gamma}\frac{{\rm d}\Gamma}{{\rm d}c_{\theta_{2}}} =
\frac{1}{2}+ \frac{1}{2}\left(-\overline{\widetilde{R}}_{21}
+\overline{\widetilde{R}}_{22}\right)c_{\theta_2}\,, \ \ \
\nonumber \\[3mm]
\frac{1}{\Gamma}\frac{{\rm d}\Gamma}{{\rm d}\Phi} &=& 
\frac{1}{2\pi}+ 
\frac{\pi}{32}\left(-\overline{\widetilde{R}}_3c_\Phi
+\overline{\widetilde{R}}_4s_\Phi \right)
\end{eqnarray}
where the decay width is given by
\footnote{In Appendix C, we show
$\Gamma=\Gamma(H\to t\bar t)\,[B(t\to b W)]^2$ in the leading order (LO)
by adopting the narrow width approximation for the intermediate top quarks.
}
\begin{equation}
\label{eq:width}
\Gamma =N_C\frac{g^4g_t^2 \sqrt{s}}{2^{12}\pi^5}
\left(\omega_{11}\overline{\widetilde{C}}_{11}+
\omega_{12}\overline{\widetilde{C}}_{12}\right)\,{\cal F}\,.
\end{equation}
When $M_H>2m_t$ and the top quarks are on-shell,
we have found that the weight factors do not deviate from
unity by more than $1\%$ and 
one may safely take $\widetilde C_i=\overline{\widetilde C_i}$ 
in Eq.~(\ref{eq:calFi}) by adopting the narrow-width approximation (NWA) 
for the intermediate top quarks.
Note that the only non-trivial one-dimensional angular distribution is 
$1/\Gamma\,{\rm d}\Gamma/{\rm d}\Phi$ since 
$\overline{\widetilde{R}}_{21,22}=0$.
Again, we note that the analytic expressions for the normalized angular 
distributions, Eq.~(\ref{eq:approx}), have been obtained in 
the $m_b=0$ limit taking $f_L=1$ and $f_R=0$.
We find that finite $m_b$ effects are negligible 
as we shall show in Fig.~\ref{fig:rp10} by comparing the distributions obtained
by using the exact expression Eq.~(\ref{eq:partial}) with finite 
$m_b$ and the expression Eq.~(\ref{eq:approx}) with $m_b=0$. We then
use Eq.~(\ref{eq:approx}) to analyze angular distributions in our numerical
analysis.

\section{Numerical Analysis}
For our numerical analysis we are taking $M_H=380$ GeV.
This choice of $M_H$ ensures 
that the top and anti-top quarks are both 
on-shell
and thus the decay channel $H\to t\bar t$ would be
the dominant decay mode of the heavy Higgs boson $H$. 
Also, the production cross section of $H$ would be substantially larger 
than the heavier Higgs boson with, for example, $M_H\gsim 500$ GeV.
For the $t$-$b$-$W$ interaction,
we assume the SM couplings: $f_L=1$ and $f_R=0$.
These input values simplify our numerical analysis and there remain
only 2 real input parameters of $g^S$ and $g^P$ to vary
\footnote{In passing, we note that
the expressions given in the last section are very general and indeed
can allow a more general analysis.}.
In this case, we find that the total decay width is given by
\begin{equation}
\label{eq:width380}
\Gamma 
\simeq 22\ g_t^2 \left[\beta_t^2\left(g^S\right)^2\ + \
       \left(g^P\right)^2\right] \ {\rm GeV}\,
\end{equation}
with ${\cal F}\simeq 6100$.  Note 
$\beta_t=0.412$ for our choice of $M_H$ close to
the $2m_t$ threshold and $m_t=173.1$ GeV
and therefore the $g^S$ contribution in the above equation
is suppressed by the $\beta_t^2$ factor.

For the $H$ couplings to top quarks,
we consider the following 6 representative scenarios:
\begin{itemize}
\item{\bf S1} : $(g^S\,,g^P) \ = \ (1\,,0)$
\item{\bf S2} : $(g^S\,,g^P) \ = \ (0\,,1)$
\item{\bf S3} : $(g^S\,,g^P) \ = \ (1\,,1)$
\item{\bf S4} : $(g^S\,,g^P) \ = \ (1\,,-1)$
\item{\bf S5} : $(g^S\,,g^P) \ = \ (1\,,0.42)$
\item{\bf S6} : $(g^S\,,g^P) \ = \ (1\,,-0.42)$
\end{itemize}
In the first two scenarios of {\bf S1} and {\bf S2}, only one of
the couplings is non-vanishing and $H$ is supposed to be a pure
CP-even (odd) state in the {\bf S1} ({\bf S2}) scenario.
In the other scenarios, CP is violated and
the couplings $g^S$ and $g^P$ take on a relative phase.
In the scenarios {\bf S5} and {\bf S6}, 
in particular, the relative sizes of the  couplings are 
chosen such that $|g^P/g^S| \sim \beta_t$ in order that
the two couplings contribute 
more or less equally to the amplitude squared.

\begin{table}
\caption{\label{tab:ratios}
{\it 
The values of $\overline{\widetilde C}_{11}+\overline{\widetilde C}_{12}$
and the 6 angular observables
$\overline{\widetilde R}_i = \overline{\widetilde C}_i/(\overline{\widetilde C}_1
+\overline{\widetilde C}_3)$
with $i=11,12,21,22,3,4$ taking $w_i=1$ 
for the 6 scenarios under consideration.
The CP and CP$\widetilde{\rm T}$ parities of each observable
are shown in the square brackets.
}
}
{\footnotesize
\begin{center}
\begin{tabular}{|c|cc|c|rrrrrr|}
\hline
& $g^S\ $ & $g^P\ $ & $(\overline{\widetilde C}_{11}+\overline{\widetilde C}_{12})[++] $ &
$\ \ \overline{\widetilde R}_{11}[++]$ & $\ \ \overline{\widetilde R}_{12}[++]$ &
$\ \ \overline{\widetilde R}_{21}[--]$ & $\ \ \overline{\widetilde R}_{22}[--]$ &
$\ \ \overline{\widetilde R}_3[++]\ \ $    & 
$\ \ \overline{\widetilde R}_4[-+]\ \ $ 
\\ \hline
{\bf S1} & $1$ & $0$ & 
$3.76$ & $0.580$ & $0.420$ &
$0$ & $0$ & $-0.159\ \ $ & $0\ \ $
\\ \hline
{\bf S2} & $0$ & $1$ & 
$22.1$ & $0.580$ & $0.420$ &
$0$ & $0$ & $0.159\ \ $ & $0\ \ $
\\ \hline
{\bf S3} & $1$ & $1$ & 
$25.9$ & $0.580$ & $0.420$ &
$0$ & $0$ & $0.113\ \ $ & $0.112\ \ $
\\ \hline
{\bf S4} & $1$ & $-1$ & 
$25.9$ & $0.580$ & $0.420$ &
$0$ & $0$ & $0.113\ \ $ & $-0.112\ \ $
\\ \hline
{\bf S5} & $1$ & $0.42$ & 
$7.67$ & $0.580$ & $0.420$ &
$0$ & $0$ & $0.00295\ \ $ & $0.159\ \ $
\\ \hline
{\bf S6} & $1$ & $-0.42$ & 
$7.67$ & $0.580$ & $0.420$ &
$0$ & $0$ & $0.00295\ \ $ & $-0.159\ \ $
\\ \hline
\end{tabular}
\end{center}
}
\end{table}
In Table \ref{tab:ratios}, we show
the values of $\overline{\widetilde C}_{11}+\overline{\widetilde C}_{12}$
and the 6 angular observables
$\overline{\widetilde R}_i = \overline{\widetilde C}_i/(\overline{\widetilde C}_1
+\overline{\widetilde C}_3)$ for the 6 scenarios under consideration
with $i=11,12,21,22,3,4$. We have taken $w_i=1$.
First of all, we observe that 
$\overline{\widetilde R}_{11}=(1+m_t^4/4M_W^4)/(1+m_t^2/2M_W^2)^2$ and
$\overline{\widetilde R}_{12}=(m_t^2/M_W^2)/(1+m_t^2/2M_W^2)^2$ independent of 
the scenario
and the CP${\tilde{\rm T}}$-odd $\overline{\widetilde R}_{21}$
and $\overline{\widetilde R}_{22}$ are identically vanishing in all the scenarios.
This leaves only $\overline{\widetilde R}_3$ and $\overline{\widetilde R}_4$
as non-trivial angular observables which can be probed 
by studying the ${\rm d}\Gamma/{\rm d}\Phi$ distribution.
The CP-odd $\overline{\widetilde R}_4$ observable is vanishing 
in the CP-conserving {\bf S1} and {\bf S2} scenarios and, if it
is not vanishing, its sign is determined by the sign of
the product of $g^S$ and $g^P$.
Further we note that $\overline{\widetilde R}_3$ is very suppressed
in {\bf S5} and {\bf S6} because it is proportional to
$-\beta_t^2\left(g^S\right)^2+\left(g^P\right)^2$.

\begin{figure}[t!]
\begin{center}
\includegraphics[width=16.3cm]{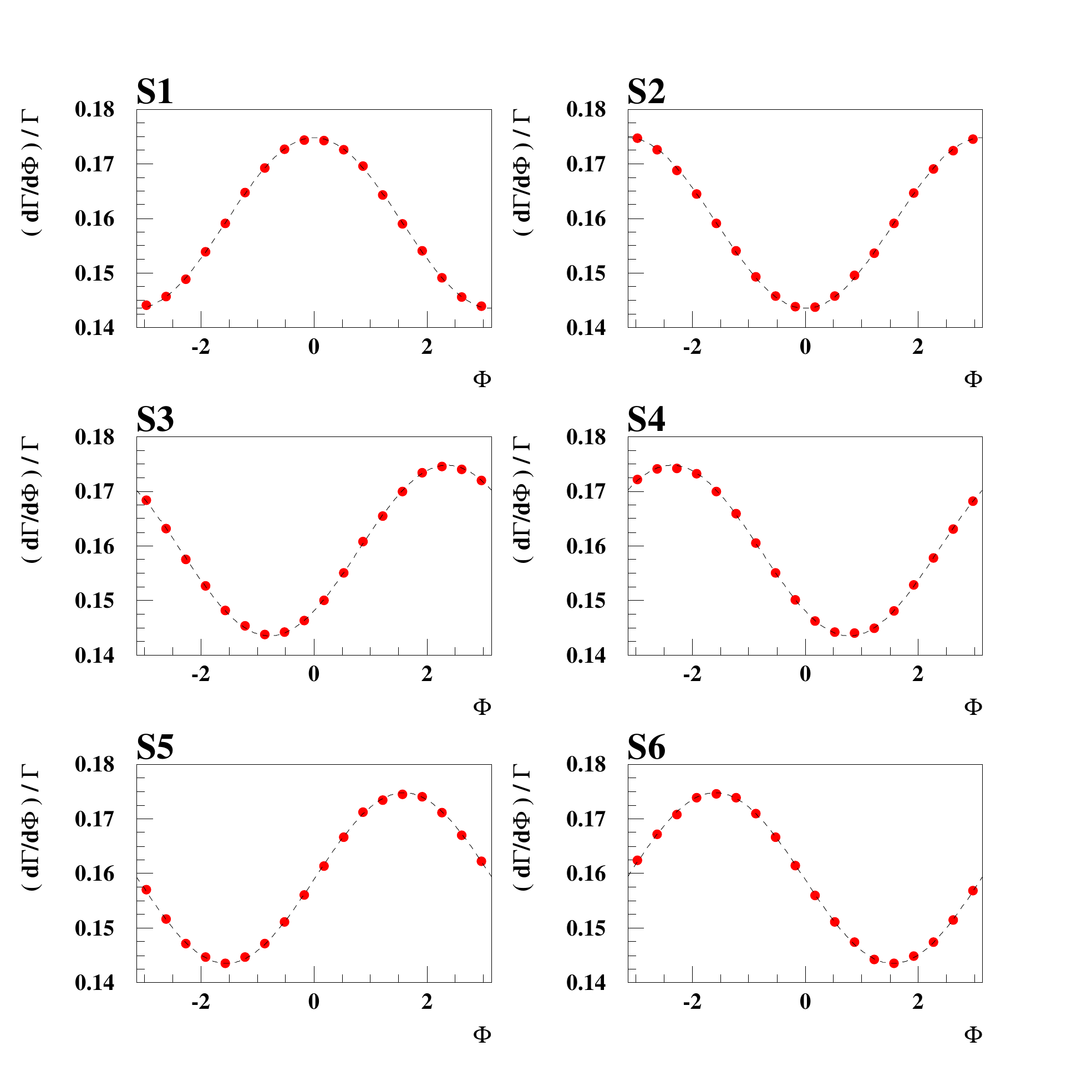}
\end{center}
\vspace{-1.0cm}
\caption{\it
The normalized angular distributions (solid dots)
generated according to Eq.~(\ref{eq:partial})
integrating over
$m_t-5\,\Gamma_t \leq \sqrt{p_t^2} \leq m_t+5\,\Gamma_t$ and
$m_t-5\,\Gamma_t \leq \sqrt{\bar p_t^2} \leq m_t+5\,\Gamma_t$.
We have taken $f_L=1$ and $f_R=0$.
In each frame, we are taking
$\left(g^S,g^P\right)=(1,0)$ ({\bf S1}: upper-left),
$\left(g^S,g^P\right)=(0,1)$ ({\bf S2}: upper-right),
$\left(g^S,g^P\right)=(1,1)$ ({\bf S3}: middle-left),
$\left(g^S,g^P\right)=(1,-1)$ ({\bf S4}: middle-right),
$\left(g^S,g^P\right)=(1,0.42)$ ({\bf S5}: lower-left), and
$\left(g^S,g^P\right)=(1,-0.42)$ ({\bf S6}: lower-right).
The dashed lines are drawn using the expressions in Eq.~(\ref{eq:approx}),
which are obtained in the $m_b=0$ limit, taking $w_i=1$.
}
\label{fig:rp10}
\end{figure}
In Fig.~\ref{fig:rp10}, we show the normalized angular distributions 
(solid dots) generated according to 
Eq.~(\ref{eq:partial}) and compare them with those 
(dashed lines)
obtained
using the analytic expressions Eq.~(\ref{eq:approx}) obtained in the $m_b=0$ limit 
taking $w_i=1$.
Without any noticeable differences between 
these two sets of distributions
in all the scenarios
\footnote{Numerically, we find that
the absolute difference is smaller than $5\times 10^{-3}$ for the 6 scenarios under
consideration.}, we conclude
that the finite $m_b$ effects are negligible and the NWA for
the angular
coefficients and observables works excellently. 
Thus, we conclude that the analytic
expressions in Eq.~(\ref{eq:approx}) provide a sufficient 
theoretical framework to analyze
the angular distributions and to extract the $g^S$ and $g^P$ couplings 
when $M_H>2m_t$.
Incidentally, the behavior of the angular distribution 
in $\Phi$ can be easily
understood as it varies according to
$-\overline{\widetilde R}_3c_\Phi+\overline{\widetilde R}_3s_\Phi$:
see Eq.~(\ref{eq:approx}) and Table \ref{tab:ratios}.

Now we are going to illustrate how well one can measure the 
properties of the 380 GeV
Higgs by taking the examples of scenarios {\bf S4} and {\bf S6}.
For this purpose, we assume that one top quark decays hadronically and the other one
leptonically with $\ell=e\,,\mu$.  Then,
the expected number of events would be
\begin{eqnarray}
\label{eq:nevt}
N_{\rm evt}&\sim&  \left(7\times 10^{3}\right) \
\left[\frac{\sigma(H)\times B(H\to t\bar t)} {10\,{\rm pb}}\right]
\left[\frac{B(t\to bW)}{1}\right]^2
\left(\frac{\epsilon_b}{0.7}\right)^2
\\[2mm] &\times &
\left[\frac{B(W^+\to {\rm hadrons})}{0.6741} \right]
\left[\frac{B(W^-\to \ell^-\bar \nu)}{0.1071+0.1063} \right]
\left(\frac{\epsilon_{\rm rec}}{0.1}\right)
\left( \frac{\cal L}{100\,{\rm fb}^{-1}}\right)  \ + \
\left(W^+ \leftrightarrow W^-\right) \nonumber
\end{eqnarray}
where $\sigma(H)$ denotes the production cross section of the heavy Higgs
boson $H$.
When $M_H\lsim 500$ GeV, the experimental constraint on
$\sigma(H)$ may come from
the search for narrow scalar resonances in the $b$-tagged dijet mass
spectrum: $\sigma(pp\to H)\cdot B(H\to b\bar b)\lsim 10$
pb at 95\% confidence level (CL)~\cite{Sirunyan:2018pas}.
Taking $B(H\to b\bar b) \sim 0.1$,
we have $\sigma(H)\lsim 100$ pb at the LHC.
Another 95\% CL limit may be derived from
$\sigma(gg\to H)\cdot B(H\to ZZ) \lsim 0.1$ pb~\cite{atlas17},
which leads to $\sigma(H)\lsim 10$ pb with
$B(H\to ZZ)\sim 0.01$.
In Eq.~(\ref{eq:nevt}),
$\epsilon_b$ denotes the $b$-tagging
efficiency and $\epsilon_{\rm rec}$ stands for a collective
efficiency of reconstructing the $H\to t\bar t$ system.
Note that $\epsilon_{\rm rec}$ includes the efficiency
of fully reconstructing
the four momenta of the $t$ and $\bar t$ quarks which are necessary to
measure the $\Phi$ distribution
\footnote{Using the pseudo-top algorithm,
for example, the missing neutrino momentum can be reconstructed
with a two-fold ambiguity at the LHC~\cite{Aad:2015eia}.
See also, for example, Ref.~\cite{Erdmann:2013rxa}
for more sophisticated algorithms for top-quark pairs.}.
We observe that $\epsilon_{\rm rec}$ may also account for
the dilutions due to interference with irreducible backgrounds
and incorrect reconstruction of the neutrino momentum.
One may achieve
$\epsilon_{\rm rec}\sim 1$ at the future $e^+e^-$ colliders but
the production cross sections would be much suppressed compared
to $pp$ colliders.
In our analysis, we are taking $N_{\rm evt}=10^4$ as reference.

\begin{figure}[t!]
\begin{center}
\includegraphics[width=8.0cm]{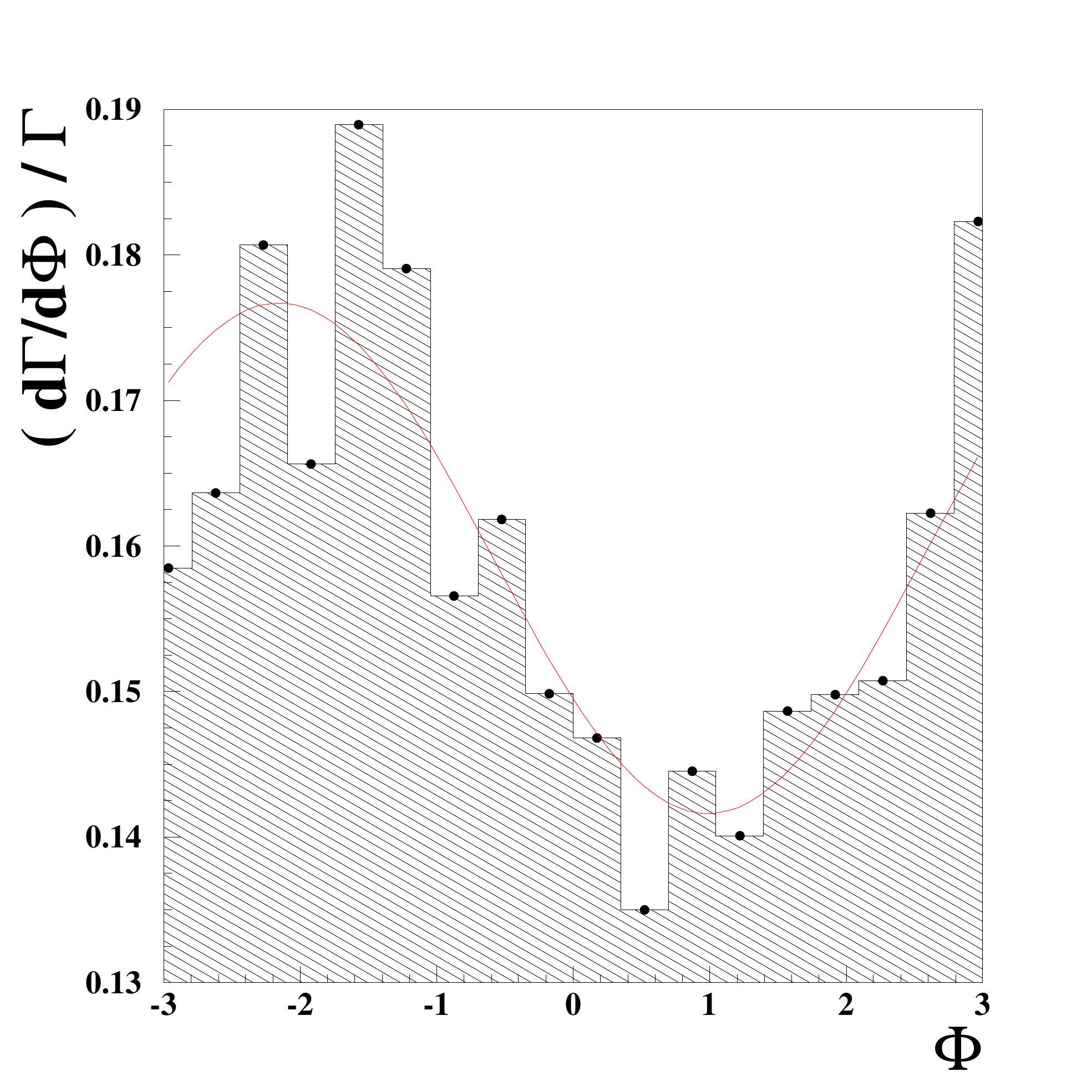}
\includegraphics[width=8.0cm]{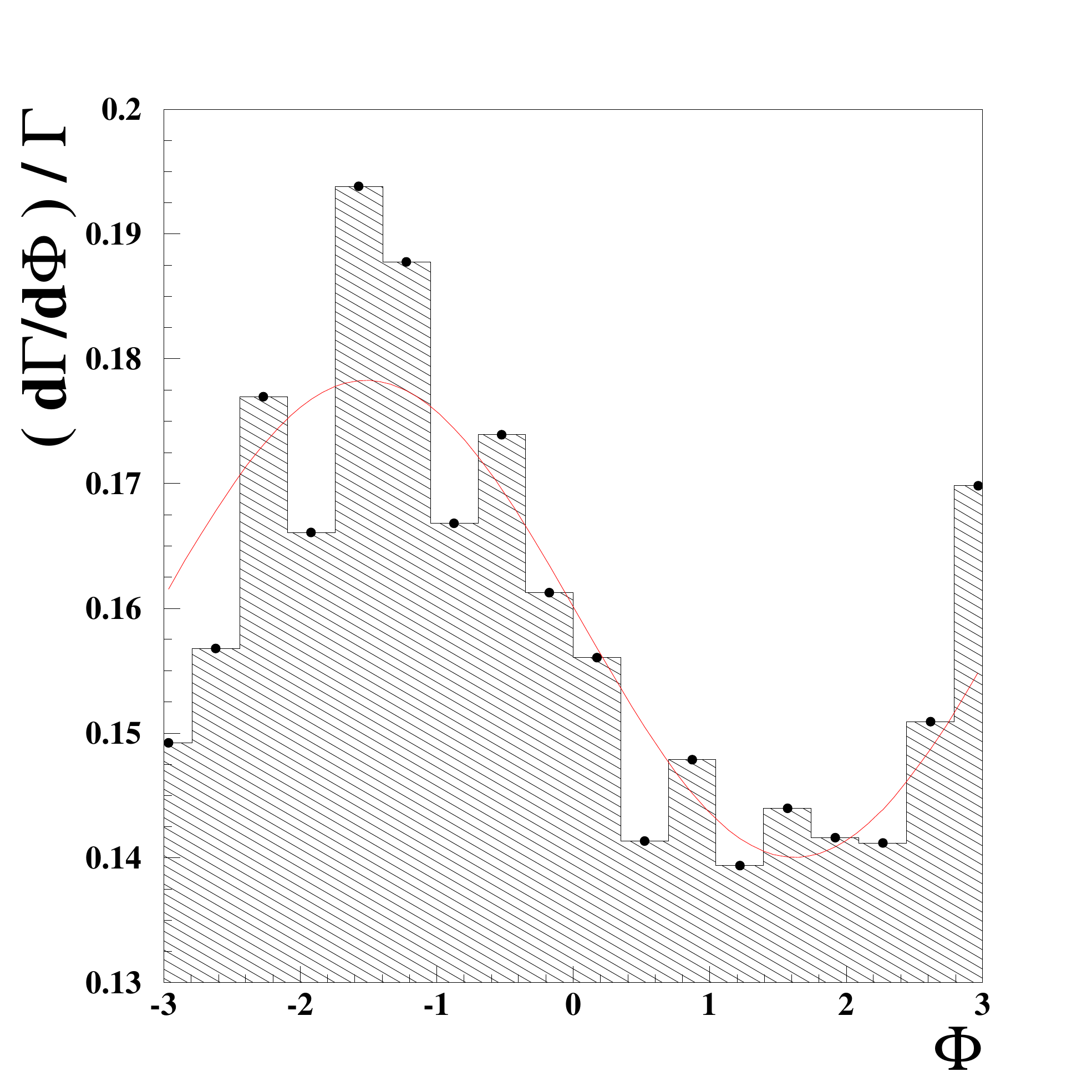}
\end{center}
\vspace{-1.0cm}
\caption{\it
{\bf  S4} (left)
{\bf  S6} (right):
The normalized angular distributions 
from the pseudo dataset generated with
$\sqrt{p_t^2}\,,\sqrt{\bar p_t^2}=m_t\pm 5 \Gamma_t$ and
$\Delta\Phi=\pi/9$.
The results of fitting to the angular
distributions with Eq.~(\ref{eq:approx}) are shown in the (red) solid lines.
}
\label{fig:phi}
\end{figure}
In Fig.~\ref{fig:phi}, we show the normalized angular distributions
for the {\bf  S4} (left) and {\bf  S6} (right) scenarios with
$N_{\rm evt}=10^4$
\footnote{In practice, we have generated 10 pseudo datasets with 
each set having $10^4$ events
and take average of them to obtain the histograms.}.
The histograms represent the pseudo-data of a total of $N_{\rm event} = 10^4$ 
generated according to Eq.~(\ref{eq:partial}).
The (red) solid lines present the results of fitting to the angular
distributions using Eq.~(\ref{eq:approx}). 
With $N_{\rm evt}=10^4$ and $\Delta\Phi=\pi/9$,
we find that the absolute size of
$1$-$\sigma$ errors of the output angular observables
of $\overline{\widetilde R}_{3,4}$ 
are about $0.02$, see Table \ref{tab:results}.
The input values are the same as in Table \ref{tab:ratios}. 
The output values together with parabolic errors have been obtained by
fitting to the normalized angular distributions 
$1/\Gamma\,{\rm d}\Gamma/{\rm d}\Phi$ in
Fig.~\ref{fig:phi}.

\begin{table}[!t]
\caption{\label{tab:results}
{\it Summary of the results obtained with $N_{\rm evt}=10^4$ and $\Delta\Phi=\pi/9$.
The input values are the same as in Table \ref{tab:ratios}. 
The output values of $\overline{\widetilde{R}}_{3,4}$
have been obtained by fitting to the normalized angular distributions in
Fig.~\ref{fig:phi}.
For $\overline{\widetilde{C}}_{11}+\overline{\widetilde{C}}_{12}$,
we simply assume $20\,\%$ error.
Implementing $\chi^2$ analysis gives the best-fit values of $g^S$ and $g^P$,
see Figs.~\ref{fig:cls_s4} and \ref{fig:cls_s6}.
Also shown are the best-fit values for
$\overline{\widetilde{C}}_{11}+\overline{\widetilde{C}}_{12}$
and $\overline{\widetilde{R}}_{3,4}$ calculated using the best-fit values of $g^S$ and
$g^P$.
}
}
\begin{center}
\begin{tabular}{|c|ccc||c|ccc|}
\hline
{\bf S4} & Input & Output & Best-fit & {\bf S6} & Input & Output & Best-fit\\[0mm]
$\chi^2_{\rm min}=0.728$ & value & value & value &
$\chi^2_{\rm min}=2.42$ & value & value & value \\[2mm] \hline
&&&&&&& \\
$\overline{\widetilde{C}}_{11}+\overline{\widetilde{C}}_{12}$ &
$25.9$ & $25.9\pm 5.18$ & $25.8$ &
$\overline{\widetilde{C}}_{11}+\overline{\widetilde{C}}_{12}$ &
$7.67$ & $7.67\pm 1.53$ & $7.73$ \\[2mm]
$\overline{\widetilde{R}}_3$ &
$+0.113$ & $+0.0985\pm 0.0228$ & $+0.0877$ &
$\overline{\widetilde{R}}_3$ &
$+0.00295$ & $-0.0102\pm 0.0228$ & $-0.00842$ \\[2mm]
$\overline{\widetilde{R}}_4$ &
$-0.112$ & $-0.149\pm 0.0229$ & $-0.133$ &
$\overline{\widetilde{R}}_4$ &
$-0.159$ & $-0.195\pm 0.0229$ & $-0.159$ \\[2mm]
$g^S$ & $+1$ & N/A & $+1.25\pm 0.22$ &
$g^S$ & $+1$ & N/A & $+1.04\pm 0.13$ \\[2mm]
$g^P$ & $-1$ & N/A & $-0.950^{+0.11}_{-0.10}$ &
$g^P$ & $-0.42$ & N/A & $-0.405\pm 0.050$  \\
&&&&&&& \\  \hline
\end{tabular}
\end{center}
\end{table}

Now we are ready to carry out a $\chi^2$ analysis to achieve
our ultimate goal of extracting the couplings $g^S$ and $g^P$ from
the angular observables $\overline{\widetilde{R}}_3$ and 
$\overline{\widetilde{R}}_4$. 
To implement the analysis, we further need 
$\overline{\widetilde{C}}_{11}+\overline{\widetilde{C}}_{12}$.
Using Eqs.~(\ref{eq:c1}) and (\ref{eq:width380}),  we have
\begin{equation}
\overline{\widetilde{C}}_{11}+\overline{\widetilde{C}}_{12}
\simeq
\frac{\left(1+\frac{m_t^2}{2M_W^2}\right)^2}{11\,g_t^2}\,
B(H\to t \bar t)\ \frac{\Gamma^H_{\rm tot}}{\rm GeV}\,.
\end{equation}
Assuming information on $B(H\to t\bar t)$ and the coupling $g_t$
can be eventually extracted from 
$\sigma\cdot B$ measurements by considering several $H$
production and decay processes, 
together with an independent measurement of the total decay width,
one may determine the combination of
$\overline{\widetilde{C}}_{11}+\overline{\widetilde{C}}_{12}$.
In our analysis, similar to the angular observables
$\overline{\widetilde{R}}_{3,4}$,
we simply assume $20\,\%$ error 
in $\overline{\widetilde{C}}_{11}+\overline{\widetilde{C}}_{12}$.


\begin{figure}[t!]
\begin{center}
\includegraphics[width=16.3cm]{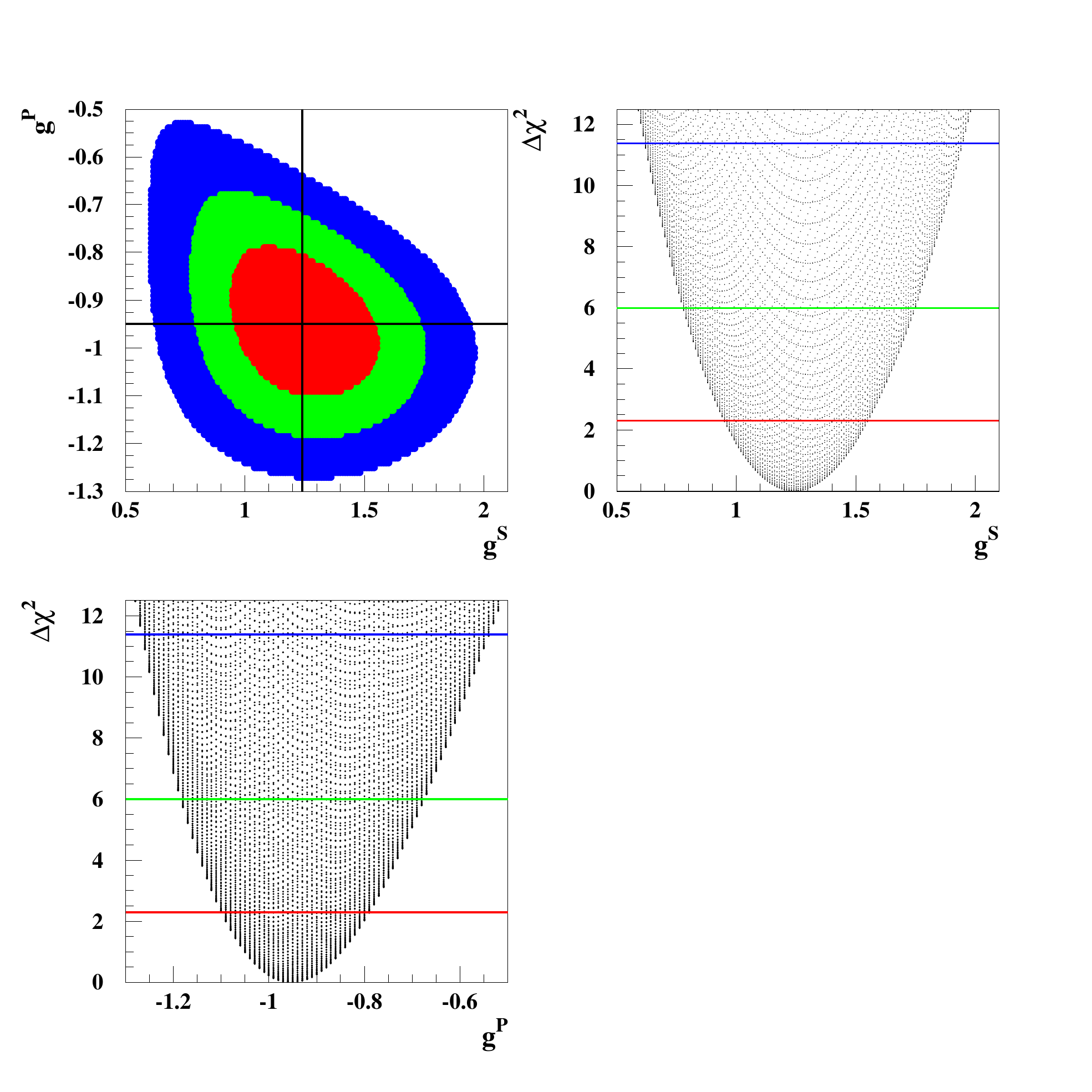}
\end{center}
\vspace{-1.0cm}
\caption{\it
Upper-left: The confidence-level (CL) regions for scenario
{\bf S4}:$\left(g^S,g^P\right)=(1,-1)$
with $\Delta \chi^2 = 2.3$ (red), $5.99$ (green), and $11.83$ (blue)
above the minimum, which
correspond to confidence levels of
$68.3\%$, $95\%$, and $99.7\%$, respectively.
The vertical and horizontal lines show the best-fit values of
$\left(g^S,g^P\right)$.
The others: The scatter plots for
$\Delta \chi^2$ versus $g^S$ (upper-right),
$\Delta \chi^2$ versus $g^P$ (lower-left).
The horizontal lines are for the $68.3\%$ (red),
$95\%$ (green), and $99.7\%$ (blue) CL regions.
}
\label{fig:cls_s4}
\end{figure}

\begin{figure}[t!]
\begin{center}
\includegraphics[width=16.3cm]{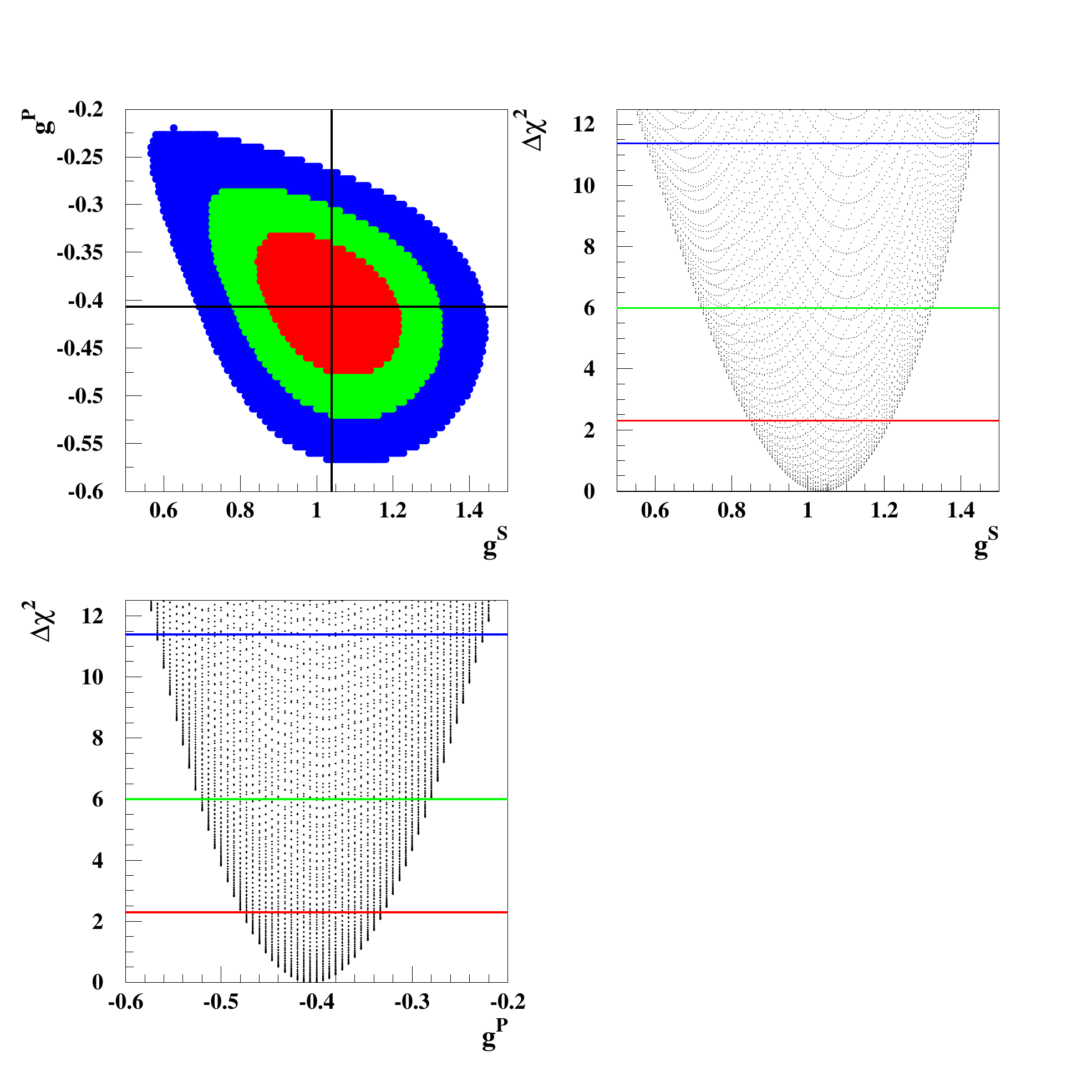}
\end{center}
\vspace{-1.0cm}
\caption{\it
The same as in Fig.~\ref{fig:cls_s4} but taking scenario
{\bf  S6}:$\left(g^S,g^P\right)=(1,-0.42)$.
}
\label{fig:cls_s6}
\end{figure}

In the upper-left frame of Fig.~\ref{fig:cls_s4},
we show the confidence-level regions of
the $\chi^2$ analysis by varying
$g^S$ and $g^P$ simultaneously for the scenario {\bf S4}.
We have found that
$\chi^2_{\rm min}/d.o.f=0.728/(3-2)=0.728$
and the minimum occurs at
\begin{eqnarray}
\label{eq:best_s4}
g^S=1.25\pm 0.22\,; \ \ \
g^P=-0.950^{+0.11}_{-0.10}\,,
\end{eqnarray}
which are consistent with the input values
$(g^S,g^P)=(+1,-1)$ within $\sim 1$-$\sigma$ range.
Therefore, we conclude that the two couplings of $H$ to the
 top-quark
pair can be determined with about
10-20\% errors when $N_{\rm evt}=10^4$ for the scenario {\bf S4}.
For the scenario {\bf S6}, the confidence-level regions 
are shown in Fig.~\ref{fig:cls_s6}. The minimum occurs at
\begin{eqnarray}
\label{eq:best_s6}
g^S=1.04\pm 0.13\,; \ \ \
g^P=-0.405\pm 0.050\,,
\end{eqnarray}
with $\chi^2_{\rm min}/d.o.f=2.42$.
We again note that the fitted values are consistent with the input values
$(g^S,g^P)=(+1,-0.42)$ safely within $1$-$\sigma$ range.
Also, we conclude that the two couplings 
can be determined with about 13\% errors in scenario {\bf S6}.
The results are also summarized in Table~\ref{tab:results}.

\section{Conclusions}

We have performed a comprehensive study of the renormalizable 
CP-even and CP-odd couplings
of a spin-0 heavy Higgs boson to a pair of top quarks,
using the angular distributions
in the decay $H\to t\bar t \to b W^+ \bar b W^-$.  Based on the
helicity amplitude method,
we figure out there are 6 types of angular observables 
$\overline{R}_i\, (i=11, 12, 21, 22, 3,4)$ 
according to their CP and CP$\widetilde{\rm T}$ parities.
We found that $\overline{R}_{21,22}$ are identically zero unless 
CP$\widetilde{\rm T}$ 
is violated through the presence of the absorptive part in the loop
correction of the $Ht\bar t$ vertex.
Furthermore, we find that among the 6 observables only the 
$\overline{R}_3$ and $\overline{R}_4$ observables
can be probed by the one-dimensional angular distribution
$d\Gamma / d \Phi$.
This is our novel strategy for analyzing 
the decay $H\to t\bar t \to b W^+ \bar W^-$
to measure the properties of a heavy Higgs boson $H$.

We have illustrated with $10^4$ events for $H\to t\bar t \to bW^+ \bar b W^-$ 
that 
the parameters $g^S$ and $g^P$
can be determined with about 10-20\% uncertainties
through the one-dimensional 
distribution $d\Gamma / d \Phi$.
This is the major numerical result of this work.

We offer the further comments in our findings:
\begin{enumerate}
\item
As long as the heavy Higgs boson is above the $t\bar t$ threshold
and, at least, as long as the angular
distributions are concerned,
the narrow-width approximation is always a good one:
the weight factors deviate from unity less
than 1\%.

\item
The angular coefficient $C_2$ 
is CP odd and CP$\widetilde{\rm T}$ odd which implies
that it
is only nonzero in presence of non-vanishing absorptive 
(or imaginary) parts of the $t \bar t H$ vertex {\it and }
in the simultaneous existence of $g^S$ and $g^P$.

\item 
The numerical analysis presented here is 
only limited to the left-handed decay of the top
quark, i.e., $f_L=1,f_R=0$. Nevertheless, the formalism here is very general
and can be applied to general studies.

\item
In the current study, while respecting the present LHC upper 
limits on $H$ production,
we have used $10^4$ events of 
$H\to t\bar t \to 2b\, 2j\, \ell\, \nu$ 
with a luminosity of 100 fb$^{-1}$.
The high-luminosity run of the LHC is supposed to 
collect 3000 fb$^{-1}$. The uncertainty in
extracting the Yukawa couplings from the heavy Higgs boson decay would
go down substantially.

\item
When the heavy Higgs boson becomes heavier, of order 1 TeV, the decay products
of the top and anti-top quarks become more boosted. In such a case, the
angular analysis is more challenging and thus deserves a new study.

\end{enumerate}

\section*{Acknowledgment}  
This work was supported by
the National Research Foundation of Korea (NRF) grant
No. NRF-2016R1E1A1A01943297.
K.C. was supported by the MoST of Taiwan under grant number 
MOST-105-2112-M-007-028-MY3.
\section*{Appendix}
\def\theequation{\Alph{section}.\arabic{equation}}
\begin{appendix}

\setcounter{equation}{0}

\section{Relation between the $t$ and $\bar t$ helicity amplitudes} 
The helicity amplitude for the decay 
$\bar t(\bar p_t,\bar \sigma_t) \to
\bar b(\bar p_b,\bar \sigma_b) W^-(k_2,\epsilon_2)$
can be obtained from that for the decay 
$t(p_t,\sigma_t) \to b(p_b,\sigma_b) W^+(k_1,\epsilon_1)$
by replacing $f_A$ with $f_{-A}^*$
together with $\sigma_{t,b} \to \bar\sigma_{t,b}$, $\lambda_1\to \lambda_2$, etc.
This can be easily understood through the relation
\begin{equation}
\label{eq:uuvv}
\bar v_t \gamma^\mu P_A v_b = \bar u_b \gamma^\mu P_{-A} u_t\,.
\end{equation}
The above relation can be shown by calculating both sides explicitly 
or by observing
\begin{equation}
C = i\gamma^2\gamma^0\,; \ \ \
C = -C^{-1} = - C^T = - C^\dagger
\end{equation}
where $C$ denotes charge conjugation and
\begin{equation}
u=C\bar v^T\,, \ \ \ \bar u = v^T C = - v^T C^{-1}\ \ ({\rm or}~ v= C\bar u^T)
\end{equation}
together with
\begin{equation}
C\left(\gamma^\mu\right)^T C^{-1} = - \gamma^\mu \,, \ \ \
C\left(\gamma^\mu\gamma_5\right)^T C^{-1} = + \gamma^\mu\gamma_5 \,. \ \ \
\end{equation}
Incidentally, $C\gamma_5^T C^{-1} = + \gamma_5$.

\section{The four-body phase space}
For the $H\to t\bar t \to bW^+\bar b W^-$ decay,
the phase space can be factorized into
\begin{eqnarray}
d\Phi_4(q\to p_t \bar p_t\to p_b k_1 \bar p_b k_2)&=&
\frac{dp_t^2}{2\pi}\
\frac{d\bar p_t^2}{2\pi}\
\frac{\lambda_H^{1/2}(1,p_t^2/s,\bar p_t^2/s)}{32\pi^2}
d\cos\Theta^*d\Phi^*\ \nonumber \\[3mm]
&\times & \frac{\lambda_t^{1/2}(1,m_b^2/p_t^2,M_W^2/p_t^2)}{32\pi^2}
d\cos\theta_1d\phi_1\ \nonumber \\[3mm]
&\times & \frac{\lambda_{\bar t}^{1/2}(1,m_b^2/\bar p_t^2,M_W^2/\bar p_t^2)}{32\pi^2}
d\cos\theta_2d\phi_2 \
\end{eqnarray}
where $s=q^2$ and $\lambda(1,a,b)=(1-a-b)^2-4ab$.
For our purpose, we may be able to take
\begin{eqnarray}
d\Phi_4(q\to p_t \bar p_t\to p_b k_1 \bar p_b k_2)&=&
\lambda_H^{1/2}\lambda_t^{1/2}\lambda_{\bar t}^{1/2}
\frac{dp_t^2}{2\pi}\
\frac{d\bar p_t^2}{2\pi}\
\frac{1}{8\pi}
\frac{d\cos\theta_1d\Phi}{32\pi^2} \
\frac{d\cos\theta_2}{16\pi} \,.
\end{eqnarray}

\section{Narrow width approximation}
Using
\begin{equation}
\delta(p^2-m^2)=\lim_{\Gamma\to 0}
\frac{m\Gamma}{\pi}\frac{1}{(p^2-m^2)^2+m^2\Gamma^2}
\end{equation}
and taking $\omega_i=1$,
we note that  Eq.~(\ref{eq:width}) together  
with Eq.~(\ref{eq:calf}) can be factorized into
\begin{equation}
\Gamma=\Gamma(H\to t\bar t)\,
\left(\frac{\Gamma^{\rm LO}(t\to b W)}{\Gamma_t}\right)^2
\end{equation}
where, taking $m_b=0$,
\begin{eqnarray}
\Gamma(H\to t\bar t)&=&
N_C\frac{\beta_t g_t^2 M_H}{8\pi}
\left[\beta_t^2\left(g_S\right)^2+\left(g_P\right)^2\right]\,, 
\nonumber \\[2mm]
\Gamma^{\rm LO}(t\to b W)&=&
\frac{g^2 m_t}{2^6\pi}\left(1-\frac{M_W^2}{m_t^2}\right)^2
\left(2+\frac{m_t^2}{M_W^2}\right) \nonumber \\ &=&
\frac{G_Fm_t^3}{8\pi\sqrt{2}}\left(1-\frac{M_W^2}{m_t^2}\right)^2
\left(1+2\frac{M_W^2}{m_t^2}\right)
\end{eqnarray}
with $G_F=\sqrt{2}g^2/8M_W^2$.
In our analysis, we are taking $\Gamma^{\rm NLO}(t\to bW)$ 
for $\Gamma_t$ or
$\Gamma_t=\Gamma^{\rm LO}(t\to bW)\,
\left[1-\frac{2\alpha_s}{3\pi}\left(\frac{2\pi^2}{3}-\frac{5}{2}\right)
\right]$ which is about $1.35$ GeV.
\end{appendix}

\clearpage


\end{document}